\documentclass[a4paper, 11pt]{article}

\usepackage[top=18truemm,bottom=25truemm,left=25truemm,right=25truemm]{geometry} 
\usepackage{indentfirst} 

\makeatletter 
\def\section{\@startsection {section}{1}{\z@}{-3.5ex plus -1ex minus -.2ex}{2.3 ex plus .2ex}  {\large\bf}}
\makeatother
\usepackage{secdot} 

\makeatletter 
\def\subsection{\@startsection {subsection}{1}{\z@}{-3.5ex plus -1ex minus -.2ex}{2.3 ex plus .2ex}{\small\bf}}
\makeatother

\usepackage{amsmath, amssymb} 

\usepackage[dvipdfmx]{graphicx}
\usepackage{float} 
\graphicspath{{./}} 
\usepackage[labelsep=period, figurename=Fig. ,tablename=Table]{caption} 
\usepackage{subcaption} 
\usepackage{booktabs}
\usepackage{multirow} 
\usepackage{tabularx}
\newcolumntype{C}{>{\centering\arraybackslash}X}

\usepackage{cite}
\usepackage{url}

\usepackage{ulem}
\usepackage{comment} 
\usepackage{color}
\usepackage{ascmac}
\usepackage{fancybox}
\usepackage{seqsplit} 

\title{Space evaluation in football games via field weighting based on tracking data}

\usepackage{authblk}
\author[1]{Takuma Narizuka\thanks{{\it E-mail address}: pararel@gmail.com (T. Narizuka).\\
}}
\author[2]{Yoshihiro Yamazaki}
\author[1]{Kenta Takizawa}
\affil[1]{Department of Physics, Faculty of Science and Engineering, Chuo University, Bunkyo, Tokyo 112-8551, Japan}
\affil[2]{Department of Physics, School of Advanced Science and Engineering, Waseda University, Shinjuku, Tokyo 169-8555, Japan}
\date{}

\begin{document}
	\maketitle

\begin{abstract}
In football game analysis, space evaluation is an important issue because it is directly related to the quality of ball passing or player formations.
Previous studies have primarily focused on a field division approach wherein a field is divided into dominant regions in which a certain player can arrive prior to any other players.
However, the field division approach is oversimplified because all locations within a region are regarded as uniform herein. 
The objective of the current study is to propose a fundamental framework for space evaluation based on field weighting.
In particular, we employed the motion model and calculated a minimum arrival time $ \tau $ for each player to all locations on the football field.
Our main contribution is that two variables $ \tau_{\textrm{of}} $ and $ \tau_{\textrm{df}} $ corresponding to the minimum arrival time for offense and defense teams are considered; using $ \tau_{\textrm{of}} $ and $ \tau_{\textrm{df}} $, new orthogonal variables $ z_{1} $ and $ z_{2} $ are defined.
In particular, based on real datasets comprising of data from 45 football games of the J1 League in 2018, we provide a detailed characterization of $ z_{1} $ and $ z_{2} $ in terms of ball passing.
By using our method, we found that $ z_{1}(\vec{x}, t) $ and $ z_{2}(\vec{x}, t) $ represent the degree of safety for a pass made to $ \vec{x} $ at $ t $ and degree of sparsity of $ \vec{x} $ at $ t $, respectively; the success probability of passes could be well-fitted using a sigmoid function.
Moreover, a new type of field division approach and evaluation of ball passing just before shoots using real game data are discussed.
\end{abstract}


\baselineskip 16pt

\section{Introduction}
The game of football, which is also commonly referred to as soccer, is a complex system wherein 22 players in two different teams interact with each other in order for their team to win.
In the process of carrying a ball to their respective goal, a vast number of movement options are available to players, leading to the emergence of various behaviors from individual to team levels, such as ball passing, dribbling, marking an opponent player, and organizing into formations.
The emergence of such behaviors in football, and recent developments in data acquisition methodologies and tools \cite{Pappalardo2019} have promoted a wide range of analyses for football games from the perspective of statistics as well as physics \cite{Sumpter2016, Gudmundsson2017}.
Examples of such analyses include the universality of goal distribution \cite{Malacarne2000}, complex network analysis for ball passing \cite{Duch2010, Buldu2019}, dynamics-based analysis of ball motion \cite{Mendes2007, Kijima2014}, and characterization of formations \cite{Bialkowski2014, Narizuka2019}.
Characterization of space in a field is a key issue in football game analysis because it is directly related to the quality of ball passing or player formations.
However, the definition of space in the context of football is not well-defined and neither are approaches for its evaluation.
In general, there are two approaches for space evaluation, namely field division and field weighting.
A well-known approach for field division is determining a ``dominant region''---as introduced by Taki et al. \cite{Taki1996, Taki2000}---wherein a certain player can arrive in a region prior to any other players.
A typical example of field division is defining a Voronoi region for each player, which corresponds to their dominant region as defined by the Euclidean distance between their position and each location in the field \cite{Okabe2000}.
The Voronoi region provides a first approximation of the territory of any player on the field, and its basic properties in football games have been previously investigated\cite{Kim2004, Fonseca2012, Ueda2014}.
However, because these Voronoi regions lack information about the velocity and acceleration of players, Fujimura and Sugihara proposed a more realistic definition of the dominant region based on a ``motion model'' \cite{Fujimura2005}.
In the motion model, each player is assumed to move according to an equation of motion with acceleration and resistance terms.
Thus, given the initial position and velocity of a player, their arrival times at any location on the field can be calculated based on the equation of motion.
Accordingly, using the equations of motion, the Voronoi regions of players can be suitably modified as per their velocity and acceleration.
Another approach for extension of the definition of a dominant region is a data-driven one wherein individual dominant regions on the field could be estimated using machine learning \cite{Gudmundsson2014, Brefeld2019}.
Though field division based on dominant regions is one approach for space evaluation when positions and motions of players are known, it is an oversimplified approach, because all locations within a dominant region yield the same space evaluation in that only one specific player can reach a location first.
For further detailed characterization of space, the field division approach could be replaced by a field weighting approach using appropriate variables.
Several recent studies have addressed the problem of field weighting based on variables such as pass probability \cite{Spearman2017}, scoring opportunity \cite{Spearman2018}, and space occupation and generation gain \cite{Fernandez2018}.
In a similar vein, in this study, we proposed yet another framework for space evaluation based on field weighting, which provides a definition of space from a new perspective.
In particular, our space evaluation approach is based on a ``minimum arrival time'' $ \tau $ to all locations in the field, which can be calculated based on the physics-based motion model.
Herein, we considered two variables $ \tau_{\textrm{of}} $ and $ \tau_{\textrm{df}} $ corresponding to the minimum arrival time for offense and defense teams, respectively.
Using $ \tau_{\textrm{of}} $ and $ \tau_{\textrm{df}} $, we introduced new orthogonal variables $ z_{1} $ and $ z_{2} $ that represent degrees of safety and sparsity of each location, respectively.
To elucidate the quantitative meaning of $ z_{1} $ and $ z_{2} $, we carried out ball-passing analyses based on real datasets.
As applications of our proposed space evaluation method, a new field division approach is discussed and an evaluation of ball passing just before shoots for a dataset of football games is presented.
%

\section{Methods}

\subsection*{Dataset and system}
The following characterization and demonstration of our space evaluation framework was based on datasets comprising data from 45 football games of the top division of the J League (i.e., Japan Professional Football League) in 2018; these datasets were provided by DataStadium Inc., Japan.
Five games were played by each of the 18 teams, and all the games were held between Aug. 10, 2018 and Sept. 2, 2018.
Each dataset contains information about the absolute positions of all players every 0.04 seconds as well as play-by-play data; the spatial resolution of the data is on the centimeter scale.
DataStadium Inc. was authorized to collect and sell this data under a contract with the J League. 
This contract also ensures that the use of relevant datasets does not infringe on any rights of the players and clubs belonging to the J League.
Though the datasets are proprietary, we received explicit permission from DataStadium Inc. for their use in this research.
The data analyses and visualizations in this study were performed using Python packages on a MacBook Pro system with a 2-GHz Intel Core i5 processor and 16 GB of memory.

\subsection*{Motion model}
Here, we summarize the motion model proposed by Fujimura and Sugihara, which is also used in this study \cite{Fujimura2005}.
In this motion model, each player is assumed to move according to the following equation of motion:
\begin{align}
	m \frac{d^{2} \vec{x}(t)}{d t^{2}} &= F \vec{n} - k \frac{d \vec{x}(t)}{d t}.
	\label{eom}
\end{align}
where $ m $ is the mass of the player and $ \vec{x}(t) $ is the position of the player at time $ t $; 
$ F $ and $ \vec{n} $ are the magnitude and direction of the attractive force, respectively; and
$ k $ is the coefficient for viscous resistance.
The solution for Eq. \eqref{eom} for an initial position $ \vec{x}_{0} $ and initial velocity $ \vec{v}_{0} $ is expressed as follows:
\begin{align}
	\vec{x}(t) &= \vec{x}_{0} + \frac{1 - \exp(-\alpha t)}{\alpha} \vec{v}_{0} + V_{\mathrm{max}} \left(t - \frac{1 - \exp(-\alpha t)}{\alpha}\right) \vec{n},
	\label{solution}
\end{align}
where $ \alpha = k/m $ and $ V_{\mathrm{max}}=F/m $ are arbitrary constants.
Thus, given $ \alpha $ and $ V_{\mathrm{max}} $, we can obtain arrival times of each player to all locations using Eq. \eqref{solution}.
In the study by Fujimura and Sugihara, they set $ \alpha=1.3 $ [1/s] and $ V_{\mathrm{max}}=7.8 $ [m/s]; these values were empirically estimated as typical values during sprinting by the player.

\section{Results}

\subsection*{Definition of variables}
\begin{figure}[H]
	\centering
	\includegraphics[width=.8\linewidth]{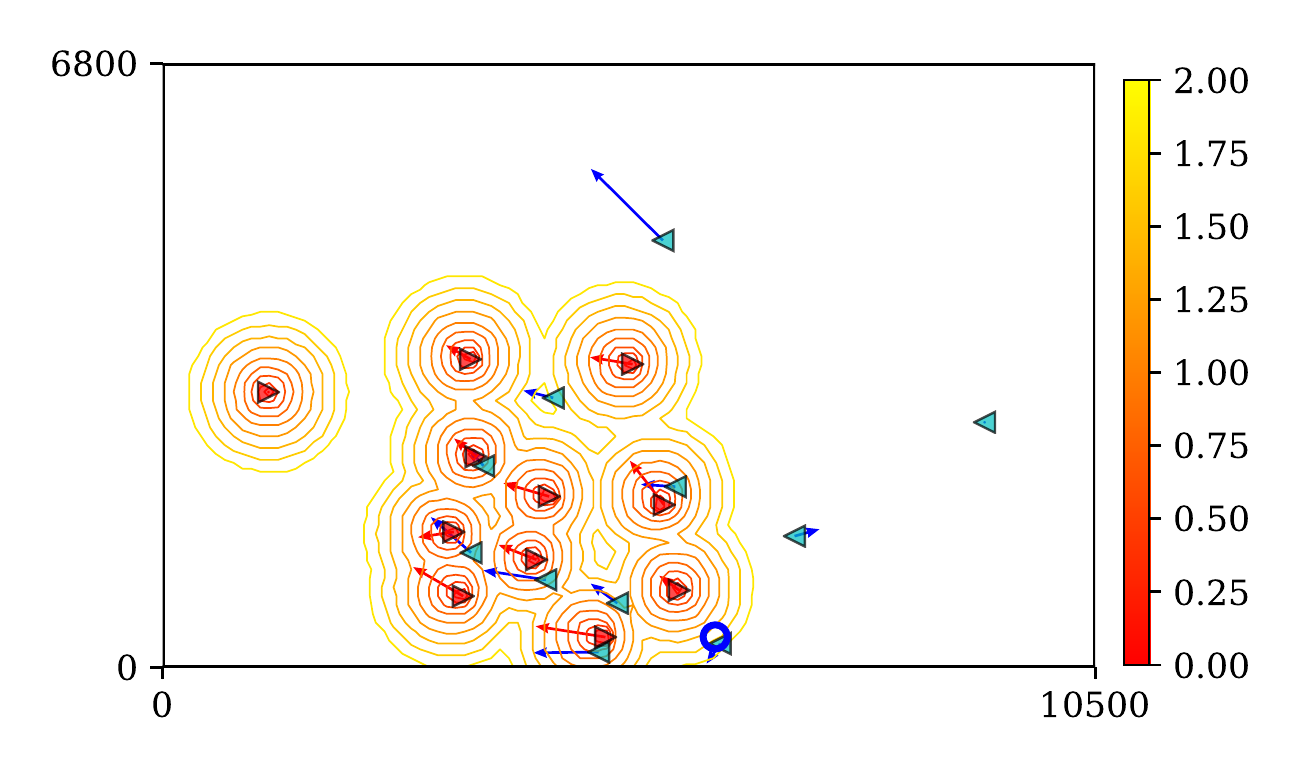}
	\caption{Visualization of $ \tau_{\textrm{df}} $ at a certain time as a contour plot over a range of 2 seconds. Players in offense (i.e., ball possession team) and defense teams are depicted using blue leftward and red rightward triangles, respectively. The blue open circle indicates the position of the ball.}
	\label{arrival_time}
\end{figure}

\begin{figure}[H]
	\centering
	\includegraphics[width=.4\linewidth]{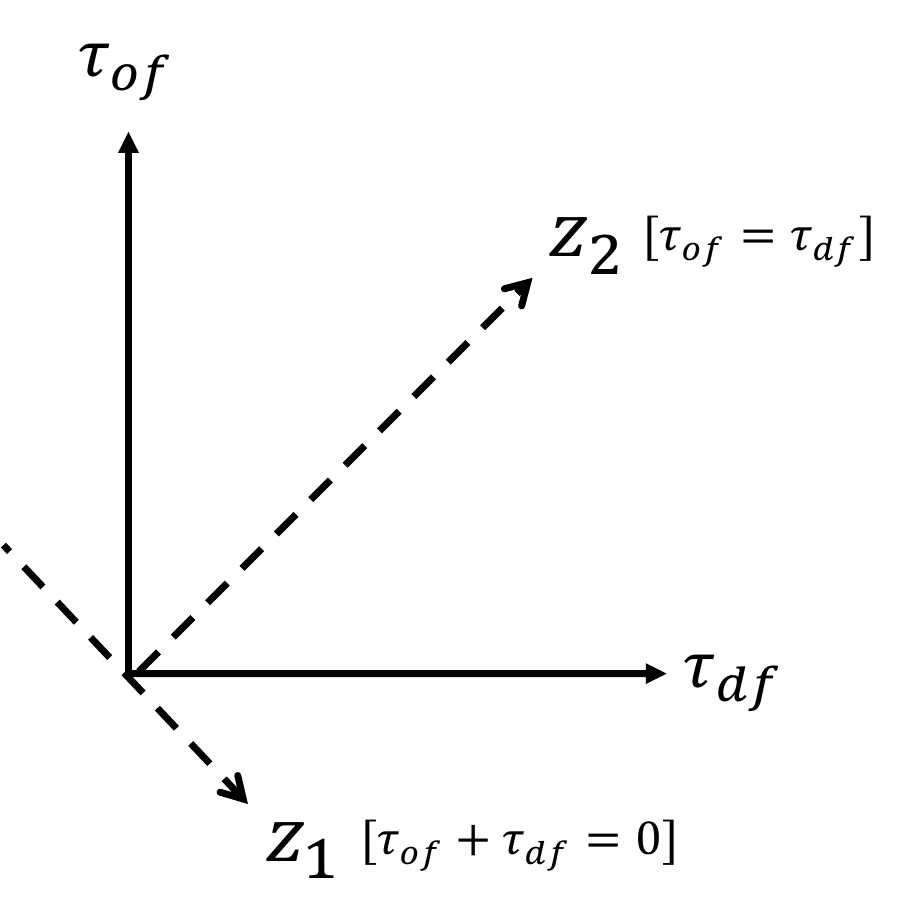}
	\caption{Relationship between the axes of $ \tau_{\mathrm{of}}$ and $\tau_{\mathrm{df}} $ and those of $ z_{1}$ and $z_{2}$.}
	\label{axes}
\end{figure}

Let $ \tau_{a}(\vec{x}, t) $ be the minimum arrival time for a player $ a $ to a location $ \vec{x} $ at time $ t $.
Accordingly, $ \tau_{a}(\vec{x}, t) $ is defined as the minimum time required for the player $ a $ to move from their position at $ t $ to $ \vec{x} $.
In order to calculate $ \tau_{a}(\vec{x}, t) $, we employed the solution of the aforementioned motion model (i.e., Eq. \eqref{solution}).
For simplicity, we set $ \alpha=1.3 $ [1/s] and $ V_{\mathrm{max}}=7.8 $ [m/s] for all players by assuming that they move to all locations by sprinting.
This simplification is considered appropriate for grasping the essential features of our space evaluation framework, i.e., the characterization of the meanings of $ z_{1} $ and $ z_{2} $.
However, for a more practical application of our framework such as assessing the playing ability of players, the use of more realistic values for $ \alpha $ and $ V_{\mathrm{max}} $ based on the play in different games is recommended.
Furthermore, we also define the minimum arrival time for a team $ A $ to $ \vec{x} $ at time $ t $ as $ \tau_{A}(\vec{x}, t) \equiv \min_{a \in A} \tau_{a}(\vec{x}, t) $.
In particular, we denote the minimum arrival time for the defense and offense teams as $ \tau_{\textrm{df}}(\vec{x}, t) $ and $ \tau_{\textrm{of}}(\vec{x}, t) $, respectively; here, the offense team is defined as the team that has possession of the ball.
Figure \ref{arrival_time} depicts a visualization for $ \tau_{\textrm{df}} $ at a certain time as a contour plot over a range of 2 seconds.
Now, we evaluate the location $ \vec{x} $ at time $ t $ using the two variables $ \tau_{\textrm{df}}(\vec{x}, t) $ and $ \tau_{\textrm{of}}(\vec{x}, t) $.
In Fig. \ref{axes}, the domain $ \tau_{\textrm{of}}(\vec{x}, t) < \tau_{\textrm{df}}(\vec{x}, t) $ corresponds to the ``safe space'' for the offense team in that an offense player can arrive in this domain prior to any defense player.
Thus, the degree of safety of $ \vec{x} $ at $ t $ for the offense team can be quantified via a signed distance $ z_{1}(\vec{x}, t) $ from the axis $ \tau_{\textrm{of}}(\vec{x}, t) = \tau_{\textrm{df}}(\vec{x}, t) $ as follows:
\begin{align}
	z_{1}(\vec{x}, t) &= \frac{\tau_{\textrm{df}}(\vec{x}, t) - \tau_{\textrm{of}}(\vec{x}, t)}{\sqrt{2}}.
	\label{z1}
\end{align}
Similarly, another variable $ z_{2}(\vec{x}, t) $, which is orthogonal to $ z_{1}(\vec{x}, t) $, can be defined as follows:
\begin{align}
	z_{2}(\vec{x}, t) &= \frac{\tau_{\textrm{df}}(\vec{x}, t) + \tau_{\textrm{of}}(\vec{x}, t)}{\sqrt{2}}.
	\label{z2}
\end{align}
Because $ z_{2}(\vec{x}, t) $ is proportional to $ \tau_{\textrm{df}} + \tau_{\textrm{of}} $, it roughly quantifies the degree of sparsity of the location $ \vec{x} $ at $ t $.
The relationship between the axes of $ \tau_{\mathrm{of}} $ and $ \tau_{\mathrm{df}} $ and those of $ z_{1} $ and $ z_{2} $ is depicted in Fig. \ref{axes}.
It is noteworthy that the definitions of $ z_{1} $ and $ z_{2} $ are independent of the definition of the motion model even though we employ the motion model proposed by Fujimura and Sugihara for one specific case in this study.
\subsection*{Calculation of $ z_{1} $ and $ z_{2} $ using ball-passing data}
For further characterization of $ z_{1} $ and  $z_{2} $ using real-world ball-passing data, we created time-series data using 34189 ball passes made in 45 football games that form part of the datasets used in our study.
A row in the datasets corresponding to a pass is expressed as $ [t_{\mathrm{o}}, \vec{x}_{\mathrm{o}}, t_{\mathrm{e}}, \vec{x}_{\mathrm{e}}, q] $.
Here, $ t_{\mathrm{o}} $ and $ \vec{x}_{\mathrm{o}} $ represent the time and positional coordinates for the origin of the pass, while $ t_{\mathrm{e}} $ and $ \vec{x}_{\mathrm{e}} $ represent those for the end of the pass, respectively.
The variable $ q \in \{1, 0\} $ indicates the success (1) or failure (0) of a pass. 
For each pass, we describe the state of the end of the pass ($ \vec{x}_{\mathrm{e}} $) at the moment the pass is made ($ t_{\mathrm{o}} $) using $ z_{1}(\vec{x}_{\mathrm{e}}, t_{\mathrm{o}}) $ and $ z_{2}(\vec{x}_{\mathrm{e}}, t_{\mathrm{o}}) $.
In order to investigate the relationship between the state of the end position and outcome of the pass (success or failure), we calculated $ z_{1}(\vec{x}_{\mathrm{e}}, t_{\mathrm{o}}) $ and $ z_{2}(\vec{x}_{\mathrm{e}}, t_{\mathrm{o}}) $ for all successful and failed passes using Eqs. \eqref{z1} and \eqref{z2}, respectively; Fig. \ref{pass_z_all}(a) shows the results of these calculations as a scatter plot.

\subsection*{Meaning of $ z_{1}(\vec{x}_{\mathrm{e}}, t_{\mathrm{o}}) $}
\begin{figure}[H]
	\centering
	\includegraphics[width=\linewidth]{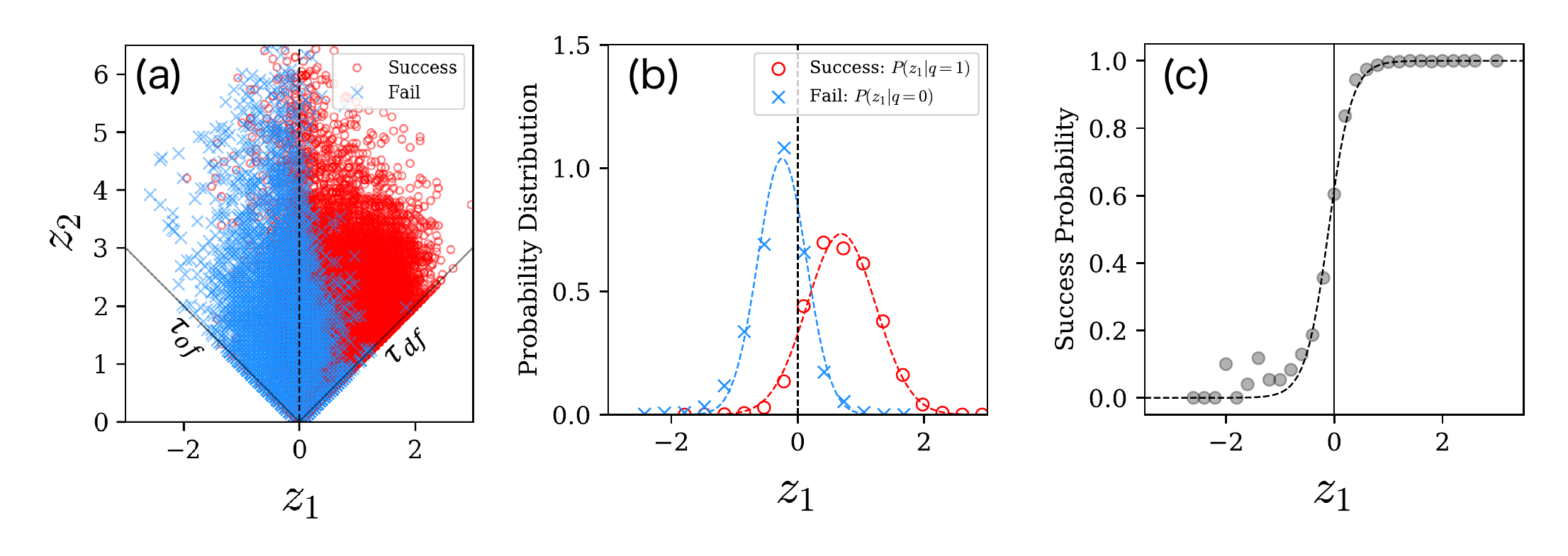}
	\caption{(a) Scatter plot of $ z_{1}(\vec{x}_{\mathrm{e}}, t_{\mathrm{o}}) $ and  $ z_{2}(\vec{x}_{\mathrm{e}}, t_{\mathrm{o}}) $ for 34189 passes obtained from 45 football games. (b) Probability distributions of $ z_{1}(\vec{x}_{\mathrm{e}}, t_{\mathrm{o}}) $ for successful and failed passes. The dotted curves are the fitted normal distribution curves. (c) Success probability of passes as a function of $ z_{1}(\vec{x}_{\mathrm{e}}, t_{\mathrm{o}}) $. The dotted curve is the fitted sigmoid function given by Eq. \eqref{sigmoid}.}
	\label{pass_z_all}
\end{figure}

Figure \ref{pass_z_all}(b) presents the probability distributions of $ z_{1}(\vec{x}_{\mathrm{e}}, t_{\mathrm{o}}) $ for successful and failed passes, which are denoted by $ P(z_{1}|q=1) $ and $ P(z_{1}|q=0) $, respectively.
From the figure, we can observe that both distributions can be fitted well using normal distribution functions.
From the corresponding normal curves, the mean and standard deviation for successful passes are obtained as $ \mu_{1}=0.69 $ and $ \sigma_{1}=0.54 $, while those for failed passes are $ \mu_{0}=-0.25 $ and $ \sigma_{0}=0.38 $.
It is noteworthy that the peak values of these normal curves--- which correspond to the mean values---are located at $ z_{1} > 0 $ and $ z_{1} < 0 $ for successful and failed passes, respectively.
The sign of $ z_{1}(\vec{x}_{\mathrm{e}}, t_{\mathrm{o}}) $ depends on whether a player of the offense or defense teams reaches the ball sooner.
Thus, it is reasonable that the sign of $ \mu $ is strongly correlated with the outcome of the pass.
Furthermore, we estimated the success probability $ P(q=1|z_{1}) $ of passes as a function of $ z_{1}(\vec{x}_{\mathrm{e}}, t_{\mathrm{o}}) $ by averaging the value of $ q $ over each $ z_{1} $.
Figure \ref{pass_z_all}(c) shows our estimation results for the success probability of passes; from these results, it was observed that these success probability values could be fitted well using a sigmoid function, which is given by:
\begin{align}
	P(q=1|z_{1}) &= \frac{1}{1 + \mathrm{exp}[-(az_{1}+b)]}
	\label{sigmoid}
\end{align}
where $ a=4.68 $ and $ b=0.48 $.
Thus, with respect to ball passing, $ z_{1}(\vec{x}, t) $ signifies the degree of safety for a pass made to $ \vec{x} $ at time $ t $.
In addition, this result suggests that the outcome of a particular pass can be estimated via logistic regression using $ z_{1} $.

\subsection*{Meaning of $ z_{2}(\vec{x}_{\mathrm{e}}, t_{\mathrm{o}}) $}

\begin{figure}[H]
	\centering
	\includegraphics[width=.8\linewidth]{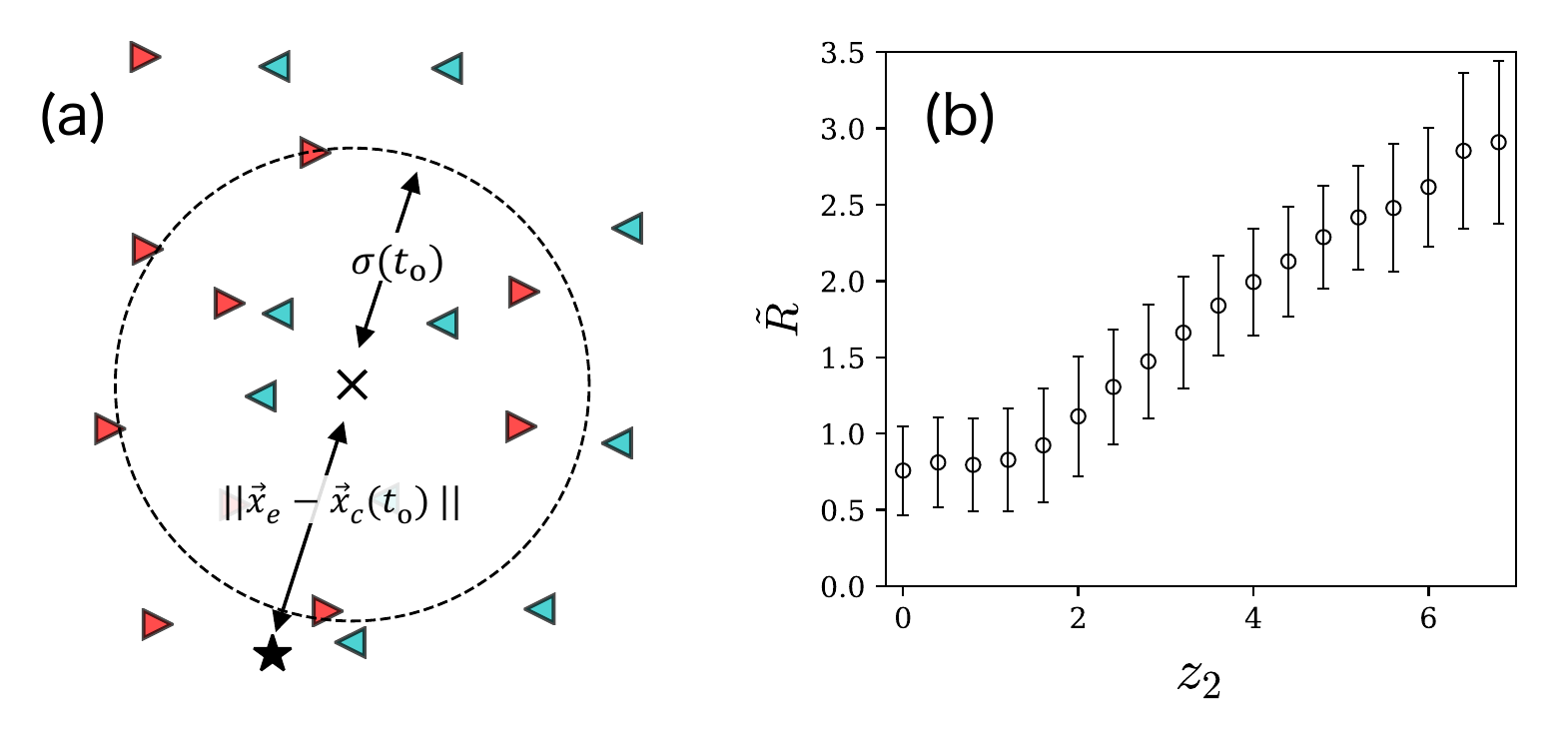}
	\caption{(a) Schematic representation of the definition of $ \tilde{R} $ (Eq. \eqref{def_R}). 
		The cross and star markers represent the end points of $ \vec{x}_{c}(t_{\mathrm{o}}) $ 
		and $ \vec{x}_{\mathrm{e}} $, respectively. 
		The dotted line indicates the circle with radius $ \sigma(t_{\mathrm{o}}) $. (b) Relationship between $ \tilde{R} $ and $ z_{2}(\vec{x}_{\mathrm{e}}, t_{\mathrm{o}}) $.}
	\label{r_z2}
\end{figure}

As mentioned earlier, $ z_{2}(\vec{x}_{\mathrm{e}}, t_{\mathrm{o}}) $ indicates the degree of sparsity of $ \vec{x}_{\mathrm{e}} $ at $ t_{\mathrm{o}} $.
To elucidate the meaning of sparsity, we calculate the following quantity for each pass:
\begin{align}
	\tilde{R}
	&= \frac{|| \vec{x}_{\mathrm{e}} - \vec{x}_{c}(t_{\mathrm{o}}) ||}{\sigma(t_{\mathrm{o}})}.
	\label{def_R}
\end{align}
Here, $ \vec{x}_{c} $ is the centroid position for all 20 field players excluding the goal keepers; $ \sigma $ is the standard deviation from $ \vec{x}_{c} $, which corresponds to the size of the formation; these variables are defined as follows:
\begin{align}
	\vec{x}_{c}(t) 
	&= \frac{1}{N}\sum_{j=1}^{20} \vec{x}_{j}(t), \\[10pt]
	\sigma(t) &= \sqrt{\frac{1}{20} \sum_{j=1}^{20}|\vec{x}_{c}(t) - \vec{x}_{j}(t)|^{2}}.
\end{align}
Figure \ref{r_z2}(a) shows the schematic representation of the definition of $ \tilde{R} $.
When $ \tilde{R} $ is less (more) than unity, the end point $ \vec{x}_{\mathrm{e}} $ of the pass is roughly located inside (outside) the formation.
The relationship between $ \tilde{R} $ and $ z_{2} $ obtained based on our data is shown in Fig. \ref{r_z2}(b).
It was observed that the value of $ z_{2} $ at which $ \tilde{R} \simeq 1 $ is $ \simeq 2 $, i.e., $ \tau_{\textrm{df}}+\tau_{\textrm{of}} \simeq 2\sqrt{2} $; thus, $ z_{2}(\vec{x}, t) \simeq 2 $ is the threshold that determines whether the location $ \vec{x} $ at $ t $ is inside or outside the formation.
Based on this result, the spaces with $ z_{2} \gtrsim 2 $ and $ z_{2} \lesssim 2 $ can be regarded as sparse and dense spaces, respectively.
%

\section{Discussion}
\begin{figure}[H]
	\centering
	\includegraphics[width=.8\linewidth]{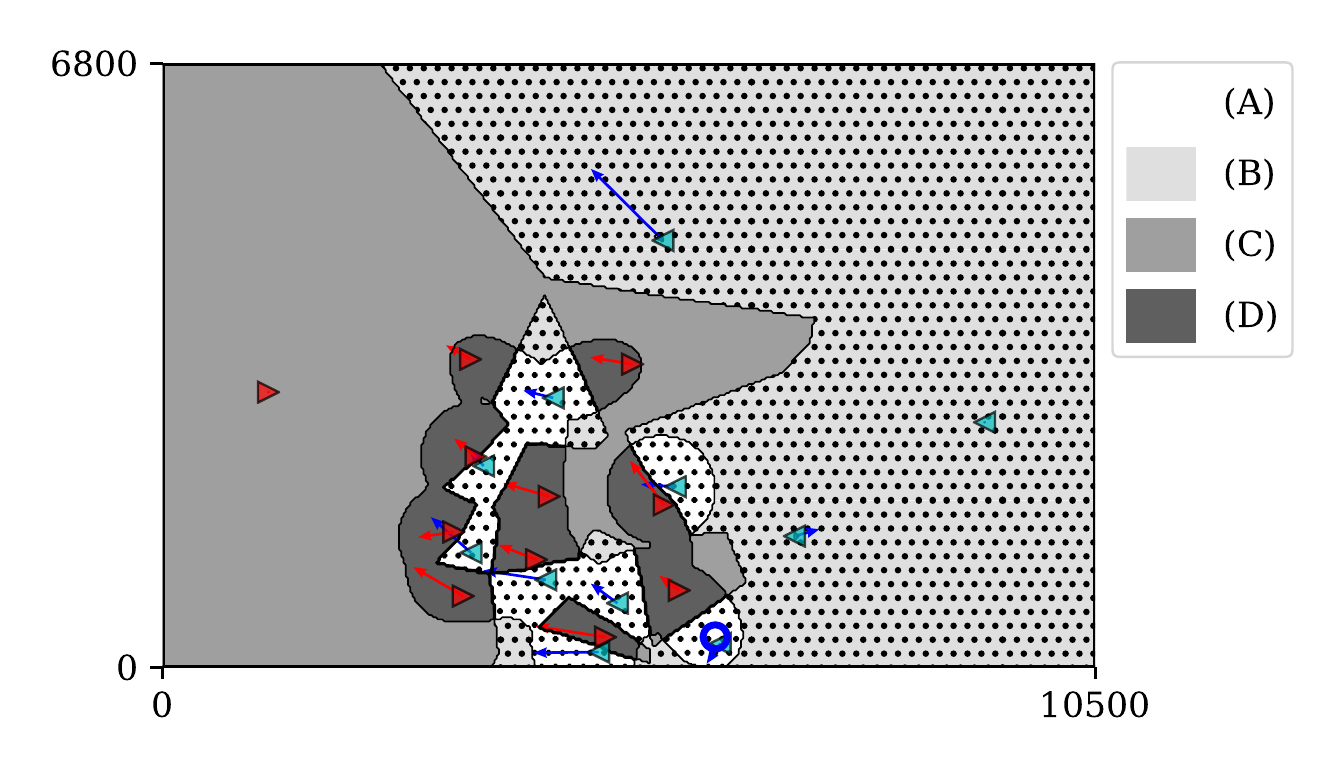}
	\caption{Typical example of field division into spaces (A)--(D). The players in the offense (i.e., ball-possession team) and defense teams are shown via blue leftward and red rightward triangles, respectively. The blue open circle indicates the position of the ball.}
	\label{division}
\end{figure}

\begin{figure}[H]
	\centering
	\includegraphics[width=.8\linewidth]{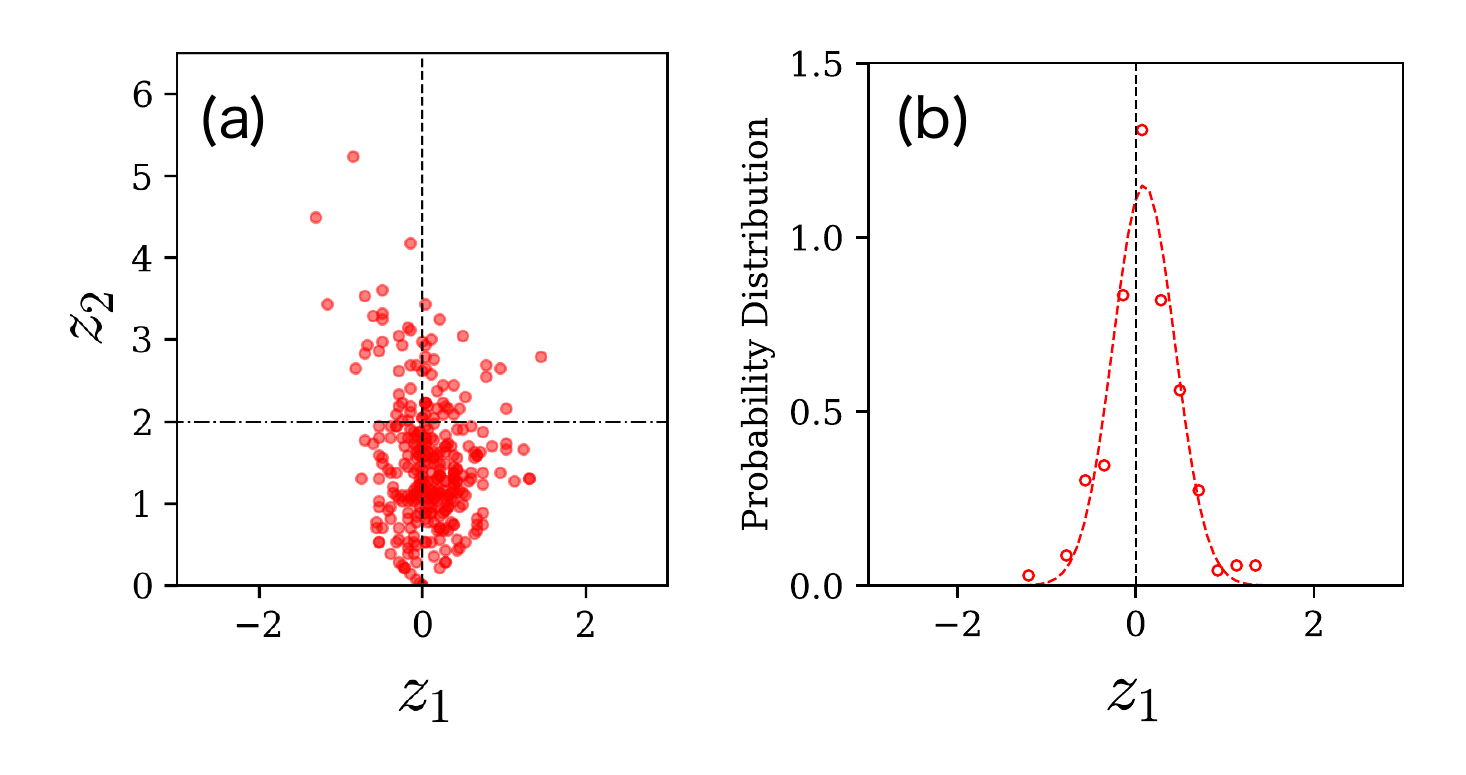}
	\caption{(a) Scatter plot of $ z_{1}(\vec{x}_{\mathrm{e}}, t_{\mathrm{o}}) $ and $ z_{2}(\vec{x}_{\mathrm{e}}, t_{\mathrm{o}}) $, and (b) probability distribution of $ z_{1}(\vec{x}_{\mathrm{e}}, t_{\mathrm{o}}) $, for successful passes just before the shoots.}
	\label{pass_z_shoot}
\end{figure}

The properties and significance of our proposed framework for space evaluation are described below.
First, instead of a field division approach, our framework is based on a field weighting approach derived from a previous motion model.
Second, the variables for field weighting have an explicit physical meaning, i.e., minimal arrival times of players when they move to all locations by sprinting.
Third, each location is evaluated using two variables; $ \tau_{\textrm{df}} $ and $ \tau_{\textrm{of}} $, and the influence of both the defense and offense teams at each location can be simultaneously evaluated using our approach.
Fourth, most importantly, two orthogonal variables $ z_{1} $ and $ z_{2} $, which correspond to the degrees of safety and sparsity of a location, were introduced.
It is important to note that our framework does not depend on the definition of a motion model; in addition, $ z_{1} $ and $ z_{2} $ provide a quantitative definition of space on the football field.
Therefore, our framework essentially yields a starting point to answer the question, "what is space in football games?"
Based on the ball-passing analysis discussed above, our framework also provides a new approach for field division based on the degrees of safety and sparsity.
In particular, using the two axes, $ z_{1}=0 $ and $ z_{2}=2 $, the football field can be divided into the following four spaces: (A) safe dense space ($ z_{1} > 0 $ and $ z_{2} < 2 $), (B) safe sparse space ($ z_{1} > 0 $ and $ z_{2} > 2 $), (C) risky sparse space ($ z_{1} < 0 $ and $ z_{2} > 2 $), and (D) risky dense space ($ z_{1} < 0 $ and $ z_{2} < 2 $).
We show a typical example of field division using our approach into spaces (A)--(D) in Fig. \ref{division}.
From the figure, it can be confirmed that the dense spaces (A) and (D) defined as $ z_{2} < 2 $ are almost located within the formation, i.e., $ \tilde{R} < 1 $.
For example, in the case shown in Fig. \ref{division}, the blue leftward triangle overlapping the blue open circle indicates that a player has the ball.
In order for this player to execute a shoot from this location, the ball has to be sent to a space in front of the opponent's goal; however, this space is risky because it corresponds to (C) or (D).
We show that passes just before shoots have a tendency to become risky generally based on our framework.
For passes just before shoots, Figs. \ref{pass_z_shoot}(a) and (b) present the scatter plots for $ z_{1}(\vec{x}_{\mathrm{e}}, t_{\mathrm{o}}) $ and  $ z_{2}(\vec{x}_{\mathrm{e}}, t_{\mathrm{o}}) $, and the probability distribution of $ z_{1} $ for successful passes, respectively.
From Fig. \ref{pass_z_shoot}(b), we can observe that the peak value $ z_{1} \simeq 0 $ is smaller than that for the case of all passes ($ z_{1} = 0.69 $ in Fig. \ref{pass_z_all}(b)).
This observation indicates that passes just before a shoot tend to become risky because the success probability of a pass is a monotonically increasing function of $ z_{1} $, as shown in Fig. \ref{pass_z_all}(c).
As shown in Fig. \ref{pass_z_all}(c), the success probability $ P(q=1|z_{1}) $ of passes being well-fitted by the sigmoid function \eqref{sigmoid} could be a consequence of the normal distributions for $ P(z_{1}|q=1) $ and $ P(z_{1}|q=0) $ (see Fig. \ref{pass_z_all}(b)).
In particular, it is known that $ P(q=1|z_{1}) $ can be expressed as follows:
\begin{align}
	P(q=1|z_{1}) &= 
	\frac{1}{1 + \exp\left[-\log \frac{P(z_{1}|q=1)P(q=1)}{P(z_{1}|q=0)P(q=0)}\right]},
\end{align}
which transforms to Eq. \eqref{sigmoid} when $ P(z_{1}|q=1) $ and $ P(z_{1}|q=0) $ follow a normal distribution with the same standard deviation \cite{Bishop2006}.
In our case, the standard deviations obtained from fitting our data are $ \sigma_{1} = 0.54 $ and $ \sigma_{0}=0.38 $, and $ \sigma_{1}\neq \sigma_{0} $.
The deviation of the success probability curve shown in Fig. \ref{pass_z_all}(c) from the sigmoid function curve where $ z_{1} < 0 $ can be attributed to this difference in standard deviation values.
Finally, here, we present directions for future research related to the present study by comparing our framework with other relevant studies.
First, Fern{\'a}ndez and Bornn proposed a field weighting approach for space evaluation in football games \cite{Fernandez2018}.
In their method, a player's influence on each location is defined by a multivariate normal distribution and is transformed into a single value to evaluate which team is dominant at a location.
In contrast, our method evaluates a location in the field based on two variables, $ z_{1} $ and $ z_{2} $, which are defined based on the motion model.
It is noteworthy that each of $ z_{1}(\vec{x}, t) $ and $ z_{2}(\vec{x}, t) $ has an explicit physical meaning, i.e., the degree of safety for a pass made to $ \vec{x} $ at $ t $ and degree of sparsity of $ \vec{x} $ at $ t $, respectively.
Several extensions to our proposed framework are required before it can be applied to real game analyses.
For example, the information of the distances from the locations of the ball and goal should be incorporated into variables for field weighting as in the case of Fern{\'a}ndez and Bornn’s approach, because the outcome of passes or shoots typically depends on these distances.
In regard to the motion of the ball, Spearman et al. proposed a model based on the equation of motion for the ball \cite{Spearman2017, Spearman2018}.
Our proposed framework could also be similarly extended to the ball, which would enable us to evaluate the accuracy of passes.
Because our framework is independent of the definition of the motion model, we could employ any extended motion model based on the equations of motion or machine learning technique \cite{Gudmundsson2014, Brefeld2019} for calculating $ z_{1} $ and $ z_{2} $.
A player-specific motion model might also allow us to assess the playing ability of players.

\section*{Acknowledgements}
The authors are grateful to DataStadium Inc., Japan, for providing the player tracking data for this study. 
This work was partially supported by the Data Centric Science Research Commons Project of the Research Organization of Information and Systems, Japan as well as by a Grant-in-Aid for Young Scientists (18K18013) from the Japan Society for the Promotion of Science (JSPS).
%

\bibliography{./reference}

\begin{thebibliography}{10}

\bibitem{Pappalardo2019}
L.~Pappalardo, P.~Cintia, A.~Rossi, E.~Massucco, P.~Ferragina, D.~Pedreschi,
  and F.~Giannotti: Scientific data {\bfseries 6} (2019) 236.

\bibitem{Sumpter2016}
D.~Sumpter: {Soccermatics: Mathematical adventures in the beautiful game}
  (Bloomsbury Sigma, London, 2016).

\bibitem{Gudmundsson2017}
J.~Gudmundsson and M.~Horton: ACM Computing Surveys {\bfseries 50} (2017).

\bibitem{Malacarne2000}
L.~Malacarne and R.~Mendes: Physica A: Statistical Mechanics and its
  Applications {\bfseries 286} (2000) 391.

\bibitem{Duch2010}
J.~Duch, J.~S. Waitzman, and L.~a.~N. Amaral: PLoS ONE {\bfseries 5} (2010)
  e10937.

\bibitem{Buldu2019}
J.~M. Buld{\'{u}}, J.~Busquets, I.~Echegoyen, and F.~Seirul.lo: Scientific
  Reports {\bfseries 9} (2019) 1.

\bibitem{Mendes2007}
R.~S. Mendes, L.~C. Malacarne, and C.~Anteneodo: The European Physical Journal
  B {\bfseries 57} (2007) 357.

\bibitem{Kijima2014}
A.~Kijima, K.~Yokoyama, H.~Shima, and Y.~Yamamoto: The European Physical
  Journal B {\bfseries 87} (2014) 41.

\bibitem{Bialkowski2014}
A.~Bialkowski, P.~Lucey, P.~Carr, Y.~Yue, S.~Sridharan, and I.~Matthews: 2014
  IEEE International Conference on Data Mining, 2014, pp. 725--730.

\bibitem{Narizuka2019}
T.~Narizuka and Y.~Yamazaki: Scientific Reports {\bfseries 9} (2019) 1.

\bibitem{Taki1996}
T.~Taki, J.-i. Hasegawa, and T.~Fukumura: Proceedings of 3rd IEEE International
  Conference on Image Processing, Vol.~3, 1996, pp. 815--818.

\bibitem{Taki2000}
T.~Taki and J.-i. Hasegawa: Proceedings Computer Graphics International 2000,
  2000, pp. 227--235.

\bibitem{Okabe2000}
A.~Okabe, B.~Boots, K.~Sugihara, and S.~Nok-Chiu: Spatial tessellations:
  concepts and applications of Voronoi diagrams (John Wiley \& Sons, New York,
  2000).

\bibitem{Kim2004}
S.~Kim: Nonlinear Analysis: Modelling and Control {\bfseries 9} (2004) 233.

\bibitem{Fonseca2012}
S.~Fonseca, J.~Milho, B.~Travassos, and D.~Ara{\'{u}}jo: Human Movement Science
  {\bfseries 31} (2012) 1652.

\bibitem{Ueda2014}
F.~Ueda, H.~Masaaki, and H.~Hiroyuki: Football Science {\bfseries 11} (2014) 1.

\bibitem{Fujimura2005}
A.~Fujimura and K.~Sugihara: Systems and Computers in Japan {\bfseries 36}
  (2005) 49.

\bibitem{Gudmundsson2014}
J.~Gudmundsson and T.~Wolle: Computers, Environment and Urban Systems
  {\bfseries 47} (2014) 16.

\bibitem{Brefeld2019}
U.~Brefeld, J.~Lasek, and S.~Mair: Machine Learning {\bfseries 108} (2019) 127.

\bibitem{Spearman2017}
W.~Spearman, A.~Basye, G.~Dick, R.~Hotovy, and P.~Pop: MIT Sloan Sports
  Analytics Conference, 2017, pp. 1--14.

\bibitem{Spearman2018}
W.~Spearman: MIT Sloan Sports Analytics Conference, 2018, pp. 1--17.

\bibitem{Fernandez2018}
J.~Fern{\'{a}}ndez and L.~Bornn: MIT Sloan Sports Analytics Conference, 2018,
  pp. 1--19.

\bibitem{Bishop2006}
C.~M. Bishop: Pattern recognition and machine learning (springer, 2006).

\end{thebibliography}

\end{document}